\documentclass[superscriptaddress,twocolumn,10pt,pra,aps,longbibliography]{revtex4-1}
\usepackage{wrapfig}
\usepackage{array}
\usepackage{graphicx}
\usepackage{amsbsy}
\usepackage{xcolor}
\usepackage[utf8x]{inputenc}
\usepackage{epstopdf}
\usepackage{amsmath,amssymb,amsfonts}

\def \sp{\rm sp}
\def \st{\rm st}

\begin{document}

\author{Ievgen I. Arkhipov}
\email{ievgen.arkhipov@gmail.com} \affiliation{RCPTM, Joint
Laboratory of Optics of Palack\'y University and Institute of
Physics of CAS, Faculty of Science, Palack\'y University, 17.
listopadu 12, 771 46 Olomouc, Czech Republic}

\author{Tai Hyun Yoon}
\affiliation{Center for Molecular Spectroscopy and Dynamics,
Institute for Basic Science (IBS), Seoul 02841, Republic of Korea}
\affiliation{Department of Physics, Korea University, Seoul 02841,
Republic of Korea}

\author{Adam Miranowicz}
\affiliation{Faculty of Physics, Adam Mickiewicz University,
PL-61-614 Poznan, Poland}

\title{Enhancing entanglement detection of quantum optical frequency combs via stimulated emission}

\begin{abstract}
We investigate the performance of a certain nonclassicality
identifier, expressed via integrated second-order intensity
moments  of optical fields, in revealing   bipartite entanglement
of quantum-optical frequency combs (QOFCs), which are generated in
both spontaneous and stimulated parametric down-conversion
processes. We show that, by utilizing that nonclassicality
identifier, one can well identify the entanglement of the QOFC
directly from the experimentally measured intensity moments
without invoking any state reconstruction techniques or homodyne
detection. Moreover, we demonstrate that  the stimulated
generation of the QOFC improves the  entanglement detection of
these fields with the nonclassicality identifier. Additionally, we
show that the nonclassicality identifier can be expressed in a
factorized form of detectors quantum efficiencies and the number
of modes, if the QOFC consists of many copies of the same two-mode
twin beam. As an example, we apply  the nonclassicality identifier
to two specific types  of QOFC, where: (i) the QOFC consists of
many independent two-mode twin beams with non-overlapped spatial
frequency modes, and (ii)  the QOFC contains  entangled spatial
frequency modes which are completely overlapped,  i.e., each mode
is entangled with all the remaining modes in the system. We show
that, in both cases, the nonclassicality identifier can reveal
bipartite entanglement of the QOFC including noise, and that it
becomes even more sensitive for the stimulated processes.
\end{abstract}

\maketitle

\section{Introduction}
One of the most striking features of quantum mechanics is quantum
entanglement~\cite{Einstein35,Schrodinger35}, which accounts for
the correlations between different parts of a system that cannot
be described within the framework of classical physics. The
development of the notion of the entanglement has led to the
establishment of new branches of physics, e.g., quantum
information theory~\cite{NielsenBook}. Apart from its theoretical
developments,  entanglement has been already experimentally tested
and exploited in quantum cryptography~\cite{Ekert1991,Gisin2002}, quantum communication~\cite{Long2002,Deng2003,Zhang2017,Chen2018,Niu2018},
quantum metrology~\cite{Giovannetti11}, quantum information
processing~\cite{Horodecki09review}, and quantum machine learning~\cite{Cai2015,Sheng2017}.

In order to implement quantum computation protocols, which utilize
quantum properties of light, one needs highly-entangled quantum
networks, e.g., continuous-variable (CV)  cluster
states~\cite{Menicucci2006}. These CV cluster states, which are
mainly Gaussian states, can be constructed from  multimode
entangled light produced, e.g., in quantum-optical frequency combs
(QOFCs)~\cite{Menicucci2008,Ukai2011,Pysher2011,Roslund2014,Cai2017}.
For quantum protocols based on discrete-variable cluster states,
see, e.g., the review~\cite{Kok2007}.

On the other hand, the problem arises how to experimentally
certify the entanglement of such multimode states. Methods have
been proposed for verifying entanglement of CV states, in
particular, Gaussian states. In most cases, these methods utilize
various nonclassicality criteria for revealing the entanglement of
such light. These include  nonclassicality criteria based on,
e.g.,  field-amplitude
moments~\cite{Hyllus2006,Loock2003,Shchukin05,Shchukin2005b,Miranowicz2006,Hillery2006,Serafini2006,Miranowicz2009,Miranowicz2010},
integrated intensity
moments~\cite{Shchukin05,Vogel2008,Arkhipov2016e,PerinaJr2017a},
anomalous-field moments~\cite{Kuhn2017}, and the measured
photocount
histograms~\cite{Haderka2005a,PerinaJr2013a,Sperling2017,arkhipov2018a},
to name a few.  Also, one may apply a CV-version of the
Peres-Horodecki entanglement criterion to the reconstructed
state~\cite{Simon00,Duan2000}, or can use entanglement witnesses
based on the separability eigenvalue
equations~\cite{Gerke2015,Gerke2018}.

In a real experiment, it is desirable to have a simple, sensitive,
and error robust tool for the entanglement identification of a
detected QOFC. One of such methods includes a simple measurement
of the mean and variance of the field intensity, which can be
performed with the help of quadratic detectors and/or
spectrometers. Of course, instead of intensities, one has to
measure the mean value and  variance of the quadratures of the
measured fields. Nevertheless, the latter seems more complicated
from the experimental point of view, as it requires balanced
homodyne detection techniques along with the use of a spatial
light modulator that has to shape the spectral profile of a local
oscillator. Thus, naturally, one would prefer to resort to some
nonclassicality identifiers (NIs) that just include the first- and
second-order intensity moments of the measured fields.

For two-mode Gaussian states generated via spontaneous parametric
processes, it has been recently shown in
Ref.~\cite{arkhipov2018c}, that with the help of stimulated
emission and by applying a certain NI, one can conclusively
identify the entanglement of such states, by measuring their
intensity moments up to the second order. We note, that a recent
study in Ref.~\cite{Kwon2019} has shown that with the measured
variances of displaced quadratures one can reveal nonclassicality
of any CV state.

In this work, motivated by the results in
Ref.~\cite{arkhipov2018c}, we study a certain NI, which is based
on the integrated second-order intensity moments, to show its
applicability in identifying bipartite entanglement of the QOFC,
which is generated in both spontaneous and stimulation parametric
down-conversion processes. In particular, we consider two
different scenarios: First, the QOFC consists of many independent
two-mode twin beams, i.e., beams with non-overlapped spatial
frequency modes. In the second scenario, the QOFC contains
completely overlapped entangled spatial frequency modes, i.e.,
each mode is entangled with all the rest modes in the system. Most
importantly, we show that with the help of the stimulation
emission, one can enhance the sensitivity of the studied NI in the
entanglement detection of the QOFC.

The paper is organized as follows. In Sec.~II we briefly review a
theory of spontaneous and stimulated down-converted QOFC. There,
we also introduce a NI, which is expressed via integrated
second-order  intensity moments, and which we use throughout
the paper. In Sec.~III, we apply the NI to the QOFC that has
non-overlapping entangled spatial frequency modes, i.e.,
independent two-mode twin beams. First,  we study the performance
of the NI for the two-mode case, and later we generalize it to any
multimode bipartitions. Additionally, we consider the effect of
stimulating fields on the performance of the NI. Section~IV is
devoted to the case when the QOFC contains completely overlapped
entangled spatial frequency modes. We show that for such QOFC, the
considered NI also proves to be useful for the identification of
multimode bipartite entanglement, and that the induced stimulation
boosts the performance of a given NI. The conclusions are drawn in
Sec.~V.

\section{Theory}
\subsection{General QOFC generated in spontaneous and stimulated down-conversion processes}

First, we briefly review  the dynamics of the quantum optical frequency comb generated
in spontaneous parametric down-conversion (PDC) process, and  that
are driven by an intense classical optical frequency comb.  Later,
we focus on the dynamics of the QOFC that are generated in the
stimulated PDC process, where the stimulating fields can also be
classical {optical frequency combs (COFCs)}. 

The dynamics of the spontaneous PDC process is described by the
following Hamiltonian in the interaction
picture~\cite{AgarwalBook,Perina1991Book}
\begin{equation}\label{H1}  
H_I=\int{\rm d}V\chi^{(2)}\hat E^{-}_p\hat E^{+}_{s}\hat
E^{+}_i+{\rm h.c.},
\end{equation}
where $\hat E_p^-$ is the negative-frequency part of the
electromagnetic field operator of the pump COFC field, and $\hat
E^{+}_{\rm s}$ ($\hat E^{+}_{\rm i}$) is the positive-frequency
part of the electromagnetic field operator of the signal (idler)
beam~\cite{AgarwalBook};  $\chi^{(2)}$ is the quadratic
susceptibility of a nonlinear medium. The integration in
Eq.~(\ref{H1}) is performed over the medium volume $V$; $\rm h.c.$
stands for Hermitian conjugate.

In the parametric approximation, the pump field, which generates
the pairs of entangled photons, can be treated classicaly. Thus,
the operator $\hat E_p$, in Eq.~(\ref{H1}), becomes a $c$-number.
For the COFC pump field, which propagates along the $z$-axis, the
electric-field amplitude can be written as follows~\cite{Lee2018}
\begin{eqnarray}\label{Ep}  
E_p(t,z)&=&\sum\limits_mA(t-m\Delta T)\exp\Big(-i\omega_p(t-m\Delta T) \nonumber \\
&&-im\Delta\phi_{ceo}+ik_{p}z\Big) \nonumber \\
&=&\exp(-i\omega_pt+ik_{p}z)\sum\limits_{n=-\infty}^{\infty}A_n\exp(-i n\omega_rt), \nonumber \\
\end{eqnarray}
where $\omega_p$ and $k_p$ are the carrier frequency and wave
vector of the pump field, respectively,  $\omega_r$ is the angular
frequency difference between the teeth of the COFC separated by the
time interval $\Delta T=2\pi/\omega_r$. The
carrier-envelope-offset phase is denoted by $\Delta\phi_{ceo}$. The
field amplitude $A_n$ corresponds to the $n$th tooth of the COFC,
i.e., to the $n$th frequency of the comb spectrum. For details
regarding COFCs, we refer the reader to the appropriate literature,
e.g., see Refs.~\cite{Hall2006,Hansch2006}.

The operators of the electric fields for both signal and idler
beams, which also propagate alone the $z$-axis, are quantized as
follows
\begin{equation}\label{Ej}  
\hat E_j^{+}=i\sum\limits_{k_j}\epsilon_{k_j}\hat
a_{k_j}\exp(-i\omega_{k_j}t+ik_jz), \quad j=s,i,
\end{equation}
where $\epsilon_{k_j}=\sqrt{2\pi\hbar\omega_{k_j}/\mu^2V}$ is the
amplitude of the electric field per photon,  $\mu$ is the
frequency-dependent refractive index, and $V$ is the quantization
volume.

Combining now Eqs.~(\ref{H1}), (\ref{Ep}), and (\ref{Ej}), and
assuming that the ideal phase-matching conditions are
satisfied~\cite{AgarwalBook,BoydBook}, i.e., $\omega_p+
n\omega_r=\omega_{k_s}+\omega_{k_i}$, and $k_p=k_s+k_i$, one
arrives at the following Hamiltonian
\begin{equation}\label{H2}  
H_I=-\hbar\sum\limits_{k_s,k_i}g_{k_s,k_i}\hat a_{k_s}\hat
a_{k_i}+{\rm h.c.},
\end{equation}
where $g_{k_s,k_i}$ is a coupling constant proportional to both
amplitude of the $n$th tooth of the COFC pump $A_n$, and the
nonlinear susceptibility $\chi^{(2)}$, and which is responsible
for the coupling between the  signal and idler modes with wave
vectors $k_s$ and $k_i$, respectively. In what follows, without
loss of generality, we assume that $g_{k_s,k_i}$ is a real-valued
parameter. The Hamiltonian in Eq.~(\ref{H2}) describes the
dynamics of the generated QOFC.

If we assume that there are $N$ different spatial frequency modes
in a QOFC, then, one can write down  the Heisenberg equations for
the boson operators, in Eq.~(\ref{H2}), as follows
\begin{equation}\label{HE2}  
\frac{{\rm d}\hat A}{{\rm d}t} = i\boldsymbol{M}\hat A,
\end{equation}
where $\hat A=(\hat a_1,\hat a_1^{\dagger},\dots,\hat a_N,\hat
a_N^{\dagger})^T$ is a $2N$-dimensional vector of the boson
annihilation and creation operators, and $\boldsymbol{M}$ is a
$2N\times2N$-dimensional evolution matrix with elements
$g_{k_s,k_i}$.

The formal  solution of Eq.~(\ref{HE2}) reads as
\begin{equation}\label{AS}   
\hat A(t) = \exp\left(i\boldsymbol{M}t\right)\hat A(0) =
\boldsymbol{S}\hat A(0),
\end{equation}
where we denoted the  matrix exponential as $\boldsymbol{S}$. Since
we consider a system with a finite number of modes, the matrix
$\boldsymbol{S}$ can always be presented as a
$2N\times2N$-dimensional matrix following the Jordan decomposition
of the matrix $\boldsymbol{M}$.

The knowledge of the quantum fields of the QOFC in Eq.~(\ref{AS})
allows one to completely characterize QOFC state through the
normally-ordered covariance matrix (CM) $\boldsymbol{A_{\cal N}}$,
which for an $N$-mode state is written as~\cite{arkhipov2018b}:
\begin{eqnarray}\label{multiCM} 
\boldsymbol{A_{\cal N}}=\begin{pmatrix}
{\bf A_1} & {\bf A_{12}} & \cdots & {\bf A_{1N}} \\
{\bf A_{12}^{\dagger}} & {\bf A_{2}} & \cdots & \vdots \\
\vdots & \vdots & \ddots & \vdots \\
{\bf A_{1N}^\dagger} & \cdots & \cdots & {\bf A_N}
\end{pmatrix}.
\end{eqnarray}
where $\bf A_k$ and $\bf A_{jl}$ are the block $2\times2$
matrices:
\begin{eqnarray}\label{lcoef}  
{\bf A_k}=\begin{pmatrix}
B_k & C_k \\
C_k^* & B_k
\end{pmatrix}, \quad
\begin{matrix}
B_k&=&\langle:\!\Delta\hat a_k^{\dagger}\Delta\hat a_k\!:\rangle, \\
C_k &=& \langle:\!\Delta\hat a_k^{2}\!:\rangle,
\end{matrix}
\end{eqnarray}
\begin{eqnarray}\label{icoef}  
{\bf A_{jl}}=\begin{pmatrix}
\bar D^*_{jl} & D_{jl} \\
D^*_{jl} & \bar D_{jl}
\end{pmatrix}, \quad
\begin{matrix}
D_{jl}&=&\langle:\!\Delta\hat a_j\Delta\hat a_l\!:\rangle, \\
\bar D_{jl}&=&\langle:\!\Delta\hat a_j^{\dagger}\Delta\hat a_l\!:\rangle, \\
\end{matrix}
\end{eqnarray}
where $\Delta \hat O = \hat O-\langle\hat O\rangle$.

To include quantum thermal noise in the system, we employ a
standard model based on the superpositions of the signal and
noise~\cite{Perina1991Book,arkhipov2015}, where the  inclusion of
this kind of noise, with the mean noise photon-number $\langle
n\rangle$, affects only the  parameters $B_{k}$  in
Eq.~(\ref{multiCM}), as $B_{k}\rightarrow B_{k}+\langle
n_{k}\rangle$, and it leaves the other elements of the
$\boldsymbol{A_{\cal N}}$ unchanged.

In the case of  stimulated PDC, i.e., when the generated QOFC is
additionally seeded by stimulating coherent fields, the dynamics
of the stimulating fields   obeys the same equation of motion as
in Eq.~(\ref{AS}) for the boson operators, namely:
\begin{equation}\label{XiS}  
\Xi(t) = \boldsymbol{S}\Xi(0),
\end{equation}
where $\Xi=(\xi_1,\xi_1^*,\dots,\xi_N,\xi_N^*)^T\in{\mathbb
C}^{2N}$ is a complex vector of $N$ stimulating  coherent fields,
and the matrix $\boldsymbol{S}$ is given in Eq.~(\ref{AS}).

With the knowledge of the covariance matrix $\boldsymbol{A_{\cal
N}}$ and the vector of stimulating coherent fields $\Xi(t)$, one
can easily arrive at the generating function $G_{\cal N}$, as
follows~\cite{arkhipov2018b}:
\begin{eqnarray} \label{GN} 
G_{\cal N}(\boldsymbol{\lambda})=\frac{1}{\sqrt{{\rm
det}\boldsymbol{ A'_{\cal
N}}}\prod\limits_{j=1}^n\lambda_j}\exp\left(-\frac{1}{2}\boldsymbol{\Xi^{\dagger}}\boldsymbol{{
A}'_{\cal N}}^{-1}\boldsymbol{\Xi}\right),
\end{eqnarray}
where $\boldsymbol{\lambda}=(\lambda_1,\dots,\lambda_n)\in{\mathbb
R}^{n}$ is a real vector. The matrix $\boldsymbol{ A'_{\cal
N}}=\boldsymbol{ A_{\cal N}}+\boldsymbol{\Lambda}^{-1}$, where the
matrix $\boldsymbol{\Lambda}^{-1}={\rm
diag}(1/\lambda_1,1/\lambda_1,\dots,1/\lambda_n,1/\lambda_n)$.

The generating function $G_{\cal N}$ allows one to obtain
statistical moments of  integrated intensities of the fields and
also their photon-number probability distribution function. The
integrated-intensity moments $ \langle W_1^{k_1}\dots
W_n^{k_n}\rangle $ are obtained from~\cite{Perina2005}:
\begin{eqnarray} \label{MOM}
\hspace{-3mm} \langle W_1^{k_1}\dots W_n^{k_n}\rangle &=& (-1)^{k_1+\dots+k_n} \nonumber \\
&&\times\left.\frac{\partial^{k_1+\dots+k_n}  G_{\cal
N}(\boldsymbol{\lambda})}{\partial\lambda_1^{k_1}\dots\partial\lambda_n^{k_n}}\right|_{\lambda_1=\dots=\lambda_n=0}.
\end{eqnarray}

\subsection{Nonclassicality identifier expressed via intensity moments}
One can write down various NIs, expressed in terms of integrated
intensity moments of the first and second order. Such an NI can be
derived either from a moments-matrix approach or, e.g., from a
majorization theory~\cite{PerinaJr2017}. At the same time, as
recent studies indicate, the moments-matrix approach enables
finding NIs with better performance than those derived from the
majorization theory~\cite{arkhipov2018a}.  Below, we present two
possible NIs based on second-order intensity moments, obtained
from the moments-matrix approach, for the entanglement
identification of  bipartite states, i.e., the entanglement
between two parts, denoted as  signal and idler arms,  as follows
\begin{equation}\label{E1}  
E_1= \langle  W_{\rm s}^{2}\rangle_{\cal N} \langle  W_{\rm
i}^2\rangle_{\cal N} - \langle  W_{\rm s} W_{\rm i}\rangle_{\cal
N}^2,
\end{equation}
and
\begin{equation}\label{E2}  
E_2= \langle \Delta W_{\rm s}^{2}\rangle_{\cal N} \langle \Delta
W_{\rm i}^2\rangle_{\cal N} - \langle \Delta W_{\rm s}\Delta
W_{\rm i}\rangle_{\cal N}^2,
\end{equation}
where $\Delta W = W-\langle W\rangle_{\cal N}$, and the moments
$\langle W_{\rm s}^{m}W_{\rm i}^{n}\rangle_{\cal N} $ are given in
Eq.~(\ref{MOM}). Whenever $E_1,E_2<0$, a bipartite state is
considered to be nonclassical, particularly, can be entangled.

One of the most important properties of the NIs, $E_1$, and $E_2$,
is that,  the quantum detection efficiencies $\eta_s$ and $\eta_i$
of the signal- and idler-beam detectors, respectively, are
factorized, i.e.,
\begin{equation}\label{E_eta}  
E_j(\eta_{\rm s},\eta_{\rm i})= \eta_{\rm s}^2\eta_{\rm i}^2E_j.
\end{equation}
Where on the r.h.s. of Eq.~(\ref{E_eta}), the NIs $E_j$ are for
the ideal case $\eta_s=\eta_i=1$. Therefore, in what follows,
without loss of generality, we always assume that one uses ideal
detectors.

As it has been already shown in Ref.~\cite{arkhipov2018b}, the NI
$E_1$ can be used for complete identification of nonclassicality
of any mixed two-mode Gaussian state by utilizing a certain
coherent displacement to the state. The NI $E_2$, as our
preliminary analysis has shown, does not possesses this
universality. Nevertheless, for particular cases, such as
multimode entangled Gaussian states, the NI $E_2$ can be even more
better than the NI $E_1$. For instance, the NI $E_2$ has a much
simpler dependence on the number of modes, that can be utilized in
more effective state reconstruction techniques based on this NI.
Moreover, in the next sections, we show that for  multimode QOFCs
with either overlapping or non-overlapping spatial frequency
modes, the NI $E_2$ demonstrates a good performance in  revealing
of  bipartite entanglement. Additionally, we show that the
stimulation of a noiseless or low noisy QOFC boosts the
performance of the NI $E_2$. In other words, the NI $E_2$
increases its negativity with the increasing intensity of
stimulating fields.  Hereafter, we denote NI $E_2$ simply as $E$.

In general, the NI $E$ can describe the nonclassicality of a
bipartite state only qualitatively not in quantitative way. In
order to relate this qualitative character of the NI $E$ to a
quantitative measure, we employ the method used in
Ref.~\cite{PerinaJr2017}. Namely, we establish the
correspondence between the NI $E$ and the Lee's nonclassicality
depth $\tau$, which is a good measure of
nonclassicality~\cite{Lee1991}. The operational meaning of $\tau$
is that it determines the amount of thermal  noise, one has to add
into both arms of a bipartite system, to remove its
nonclassicality. When considering a two-mode state, we relate
$\tau$ to the least negative eigenvalue, taken with opposite sign,
of the two-mode covariance matrix $\boldsymbol{A_{\cal
N}}$~\cite{arkhipov2016b}. In this case, the condition $\tau>0$ is
both necessary and sufficient for the nonclassicality of the
two-mode state.  For the multimode case, when considering
multimode bipartitions involving $M$ modes, we refer to $\tau_M$
as the $\tau$-parametrized NI, $E_{\tau_M}$, that can be written
as follows~\cite{PerinaJr2017}:
\begin{eqnarray}\label{tauM}  
E_{\tau_M}&=&\tau_M^4+2\tau_M^3\Big(\langle W_{\rm s}\rangle_{\cal N}+\langle W_{\rm i}\rangle_{\cal N}\Big) \nonumber \\
&&+\tau_M^2\Big(\langle \Delta W_{\rm s}^{2}\rangle_{\cal N}+\langle \Delta W_{\rm i}^{2}\rangle_{\cal N}+4\langle W_{\rm s}\rangle_{\cal N}\langle W_{\rm i}\rangle_{\cal N}\Big) \nonumber \\
&&+2\tau_M\Big(\langle \Delta W_{\rm s}^{2}\rangle_{\cal N}\langle W_{\rm i}\rangle_{\cal N}+\langle \Delta W_{\rm i}^{2}\rangle_{\cal N}\langle W_{\rm s}\rangle_{\cal N}\Big)+E_M, \nonumber \\
\end{eqnarray}
which  determines the amount of thermal noise $\tau_M$, that one
also has to add into both signal and idler arms of a bipartite
$M$-mode state to erase the negativity of  $E_M$. In other words,
the amount of nonclassicality $\tau_M$ is defined from the
condition $E_{\tau_M}=0$. Importantly, in this case, the condition
$\tau_M>0$ is no more necessary but only sufficient for the
nonclassicality of a bipartite QOFC state. Since  $\tau_M>0$ holds
only when $E_M<0$, according to Eq.~(\ref{tauM}). But the
condition $E_M<0$ is sufficient but not necessary for the
nonclassicality identification. The reason why we resort to the
$\tau_M$, derived from Eq.~(\ref{tauM}), and not from the
multimode covariance matrix $\boldsymbol{A_{\cal N}}$, is that for
a large number of modes, $M\gg1$, the problem of finding the
eigenvalues of a large-size matrix becomes computationally hard.
Nevertheless,   $\tau_M$ can serve as a nonclassicality quantifier
for bipartite states of the QOFC.

\section{Entanglement identification of QOFC with spatially non-overlapping frequency modes}
In this section, we apply the NI $E$, given in Eq.~(\ref{E2}), to
the QOFC that consists of non-overlapping spatial frequency modes,
i.e., any spatial frequency mode $k_s$ is entangled with only one
given mode $k_i$. In other words, this QOFC is comprised of many
independent two-mode twin beams.
 We note that,  in general, the down-converted frequency modes constituting QOFC,  which are generated
by different frequencies of the COFC pump, can overlap. The latter
case is considered in the next section. Here, instead, we consider
the case when such overlapping does not occur.  Such  QOFC has
already been experimentally realized in Ref.~\cite{Avella2014},
and, for example, in cavity-enhanced spontaneous
PDC~\cite{Ou1999,Kuklewicz2006,Nielsen2007,Wang2008,Scholz2009}.
Additionally, to make our  analysis simpler, we will first focus
on a two-mode case and, then, we will proceed to the multimode
scenario.

\subsection{Two-mode entanglement}
\subsubsection{Spontaneous PDC}
\begin{figure}[t!] 
\includegraphics[width=0.43\textwidth]{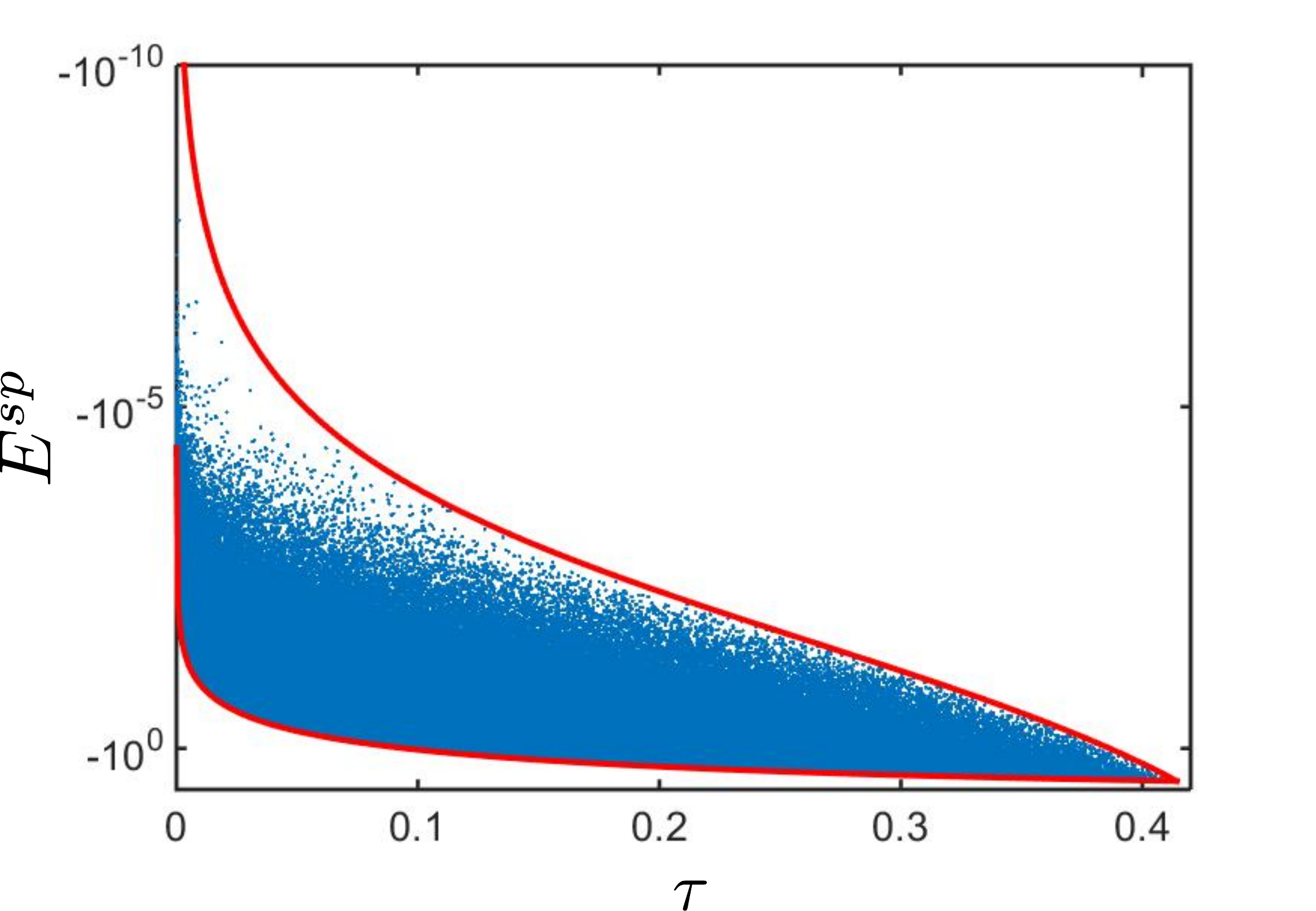}
\caption{Nonclassicality identifier $E^{\sp}$, given in
Eq.~(\ref{En}), versus the Lee's nonclassicality depth $\tau$ for
randomly generated $10^6$ states of a mixed two-mode twin beam
using a Monte-Carlo simulation. Each point on the graph denotes a
certain mixed twin beam state for which $E^{\sp}$  and $\tau$ are
calculated. Upper and lower red solid curve intersect the line
$\tau=0$ only at the point $E^{\sp}=0$.}\label{fig1}
\end{figure}
For the QOFC, with non-overlapping spatial frequency modes, which
is generated in a spontaneous PDC, the boson operators of the
signal and idler modes of a two-mode twin beam, produced by the
$n$th tooth of the COFC pump,  according to Eq.~(\ref{AS}),
acquires the following form
\begin{eqnarray}\label{FO}  
\hat a_{{\rm s},n}(t) &=& \cosh (g_nt)\hat a_{{\rm s},n}(0)+i\sinh (g_nt)\hat a_{{\rm i},n}^{\dagger}(0), \nonumber \\
\hat a_{{\rm i},n}(t) &=& \cosh (g_nt)\hat a_{{\rm i},n}(0)+i\sinh
(g_nt)\hat a_{{\rm s},n}^{\dagger}(0).
\end{eqnarray}
For simplicity, we drop the subscript $n$ in the boson operators.

In that case, the covariance matrix $\boldsymbol{A_{\cal N}}$, in
Eq.~(\ref{multiCM}), of the whole QOFC is factorized on a set of
independent $4\times4$ matrices, each corresponding to a two-mode
twin beam. The nonzero elements of a given two-mode covariance
matrix read as follows:
\begin{equation}\label{el_sp_1}  
B_j = B_{\rm p}+\langle n_j\rangle, \quad D_{\rm si}=i\sqrt{B_{\rm
p}(B_{\rm p}+1)}, \quad j=\rm s,i,
\end{equation}
where $B_{\rm p}=\sinh^2gt$ is the mean photon number of entangled
pairs, and $\langle n_j\rangle$ is the mean thermal noise
photon-number in $j$th mode.

Combining now together Eqs.~(\ref{el_sp_1}), (\ref{GN}), and
(\ref{MOM}), the  NI $E$, in Eq.~(\ref{E2}), can be written as
\begin{equation}\label{En}  
E^{\sp}=\left(B_{\rm s}B_{\rm i}-|D_{\rm si}|^2\right)\left(B_{\rm
s}B_{\rm i}+|D_{\rm si}|^2\right),
\end{equation}
where superscript $\sp$ in $E^{\sp}$ accounts for spontaneous PDC.

The expression in the first bracket, in Eq.~(\ref{En}), is a
Fourier determinant of the normally-ordered characteristic
function  of the two-mode twin beam~\cite{Perina2005}. Hence, when
this determinant is negative, the Glauber-Sudarshan $P$ function,
which is the Fourier transform of  the normally-ordered
characteristic function, fails to be a classical distribution
function~\cite{Perina1991Book,AgarwalBook}. The latter serves as a
definition of the nonclassicality and, therefore, determines the
entanglement of the twin beam state. Therefore, whenever a twin
beam is entangled,  $E^{\sp}$ always attains negative values. As
such, $E^{\sp}$ becomes a genuine NI for the two-mode twin beams.
Figure~\ref{fig1} shows the dependence of the NI $E^{\sp}$ on the
Lee's nonclassicality depth $\tau$.   This graph indicates that
$E^{\sp}$ is  a nonclassicality monotone for any mixed two-mode
twin beam, i.e., whenever $\tau>0$, then $E^{\sp}<0$.

For pure two-mode twin beams, the NI $E^{\sp}$ attains a simple
form
\begin{equation}  
E^{\sp}=-B_{\rm p}^2(2B_{\rm p}+1).
\end{equation}
Hence, the more intense is the twin beam the larger is its
entanglement and, thus, the greater is the negativity of
$E^{\sp}$.
\begin{figure} 
\includegraphics[width=0.43\textwidth]{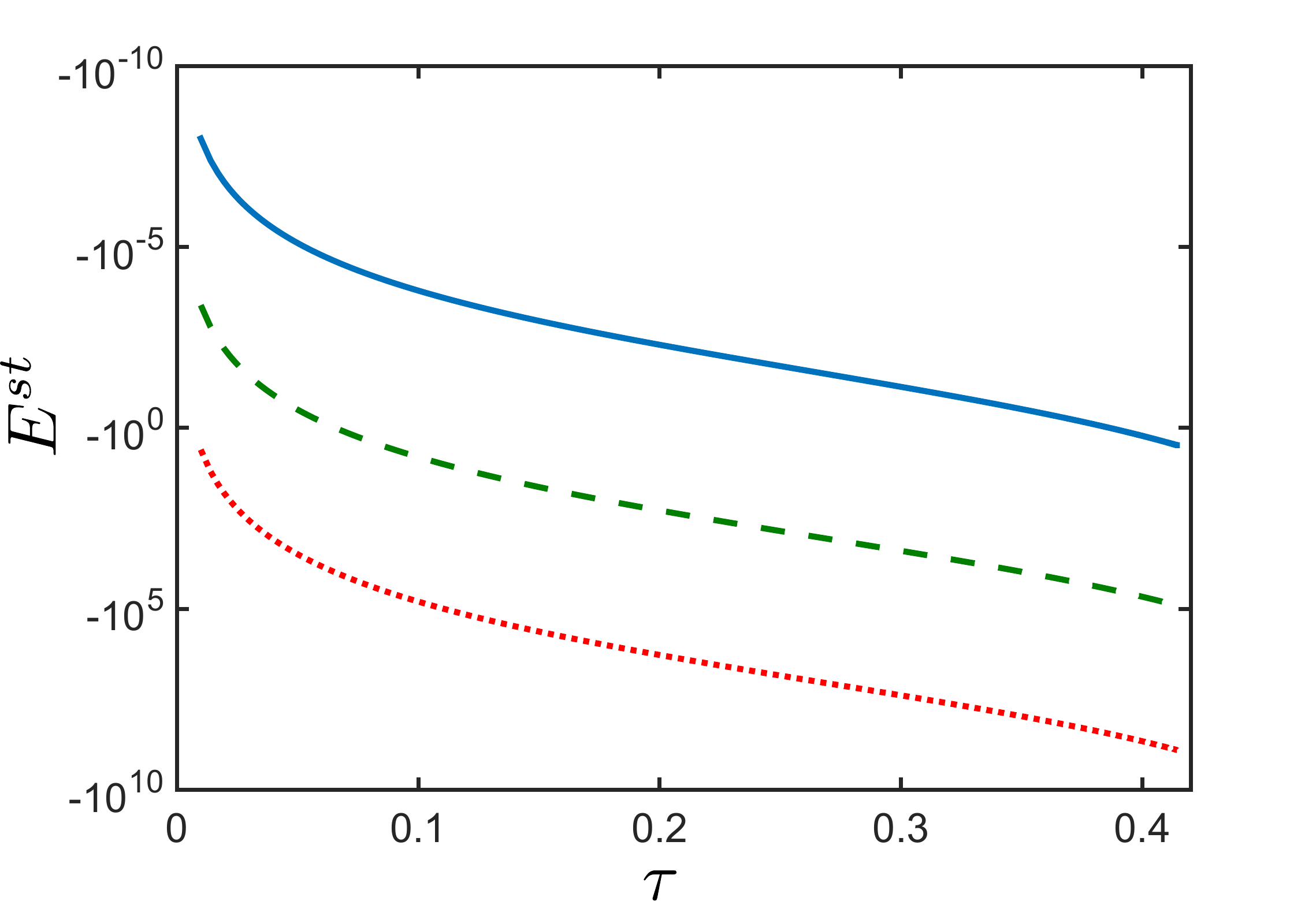}
\caption{Nonclassicality identifier $E^{\st}$,  according to
Eq.~(\ref{E_st}), versus the nonclassicality depth  $\tau$ for
pure stimulated twin beams. The stimulation is applied only in the
signal field. The NI $E^{\st}$ when the stimulating signal field
is: $\xi_{\rm s}=0$ (blue solid curve),   $\xi_{\rm s}=10$ (green
dashed curve), and  $\xi_{\rm s}=100$ (red dotted curve). The mean
photon-number of pairs $B_{\rm p}\in(0,1]$. For a given value of
$\tau$, by increasing the intensity of the stimulating field, the
negativity of $E^{\st}$ also increases. }\label{fig2}
\end{figure}


\subsubsection{Stimulated PDC}
In a stimulated PDC process, the generated twin beam at the output
of a nonlinear crystal contains a nonzero coherent part due to the
presence of  stimulating coherent fields. The stimulation process
of the twin beams, generated by a COFC pump, can be realized by
another COFC that seeds both signal and idler fields. The dynamics
of such stimulating fields, which stimulate the $n$th twin beam,
as given in Eq.~(\ref{FO}), reads according to Eq.~(\ref{XiS}), as
follows
\begin{eqnarray}\label{xi}  
\xi_{{\rm s}}(t)&=&\cosh (gt)\xi_{{\rm s}}(0)+i\sinh (gt)\xi_{{\rm i}}^*(0), \nonumber \\
\xi_{{\rm i}}(t)&=&\cosh (gt)\xi_{{\rm i}}(0)+i\sinh (gt)\xi_{{\rm
s}}^*(0).
\end{eqnarray}
Hereafter, for simplicity, we assume that the stimulation process is
performed by a seeding COFC that stimulates only the signal field,
i.e., $\xi_{\rm i}(0)=0$.

For pure states,  the NI $E$, then, acquires the following form
\begin{eqnarray}\label{E_st}  
E^{\st}=&&-B_{\rm p}^2(2B_{\rm p}+1) \nonumber \\
&&- 4B_{\rm p}^2|\xi_{\rm s}(0)|^2\left[|\xi_{\rm s}(0)|^2(B_{\rm p}+1)+\frac{3}{2}B_{\rm p}+1\right], \nonumber \\
\end{eqnarray}
where the first term accounts for the negativity of $E^{\st}$ due
to spontaneous emission, and the second term corresponds to
stimulated emission. For a given value of $\tau$, $E^{\st}$
increases its negative value with the increasing amplitude of
stimulating field $\xi_{\rm s}$ (see Fig.~\ref{fig2}). This means
that, the stronger is the  stimulating field $\xi_{\rm s}$, the
more  negative is $E^{\st}$. Moreover, as indicated by
Eq.~(\ref{E_st}),  $E^{\st}$ is independent of the phase of the
 stimulating field  $\xi_{\rm s}$ and it depends solely on the
coherent field intensity.


\subsection{Multimode bipartite entanglement}
Now, we apply the NI $E$, as denoted by $E_M$, for certifying
bipartite entanglement of the multimode QOFC,  consisting of $M$
independent two-mode twin beams. By performing bipartition of a
multimode twin beam such that all the signal modes belong to the
signal arm, and all the idler modes to the idler arm, $E_M$, then,
can be written as follows
\begin{eqnarray}\label{E_M}  
E_M=&&\sum\limits_{n=1}^M\langle \Delta W_{{\rm s},n}^{2}\rangle_{\cal N}\sum\limits_{n=1}^M\langle \Delta W_{{\rm i},n}^{2}\rangle_{\cal N} \nonumber \\
&&- \left(\sum\limits_{n=1}^M\langle \Delta W_{{\rm s},n}\Delta
W_{{\rm i},n}\rangle_{\cal N}\right)^2,
\end{eqnarray}
where
\begin{eqnarray}\label{WsWi}  
\langle \Delta W_{a,n}^{2}\rangle_{\cal N}&&=B_{a,n}\left(B_{a,n}+2|\xi_{a,n}(t)|^2\right), \nonumber \\
\langle \Delta W_{{\rm s},n}\Delta W_{{\rm i},n}\rangle_{\cal N}& &=-2{\rm Re}\left[\xi_{{\rm s},n}(t)\xi_{{\rm i},n}(t)D_{{\rm si},n}^*\right]-|D_{{\rm si},n}|^2. \nonumber \\
&&a={\rm s,i}, \quad n=1,\dots,M.
\end{eqnarray}
Note that  we have assumed a general stimulated PDC process in the
derivation of Eq.~(\ref{WsWi}).

If the system is comprised of
 $M$ copies of the same two-mode twin beam with the same stimulation, one then attains
\begin{equation}\label{E_Mc}  
E_M=M^2E,
\end{equation}
where $E$ is the NI for a two-mode twin beam copy, which we
considered earlier. Thus, the number $M$ of copies of the same
two-mode twin beam serves as a coherent multiplier of the
negativity of  $E_M$.

\subsubsection{Spontaneous PDC}
\begin{figure}[t!] 
\includegraphics[width=0.48\textwidth]{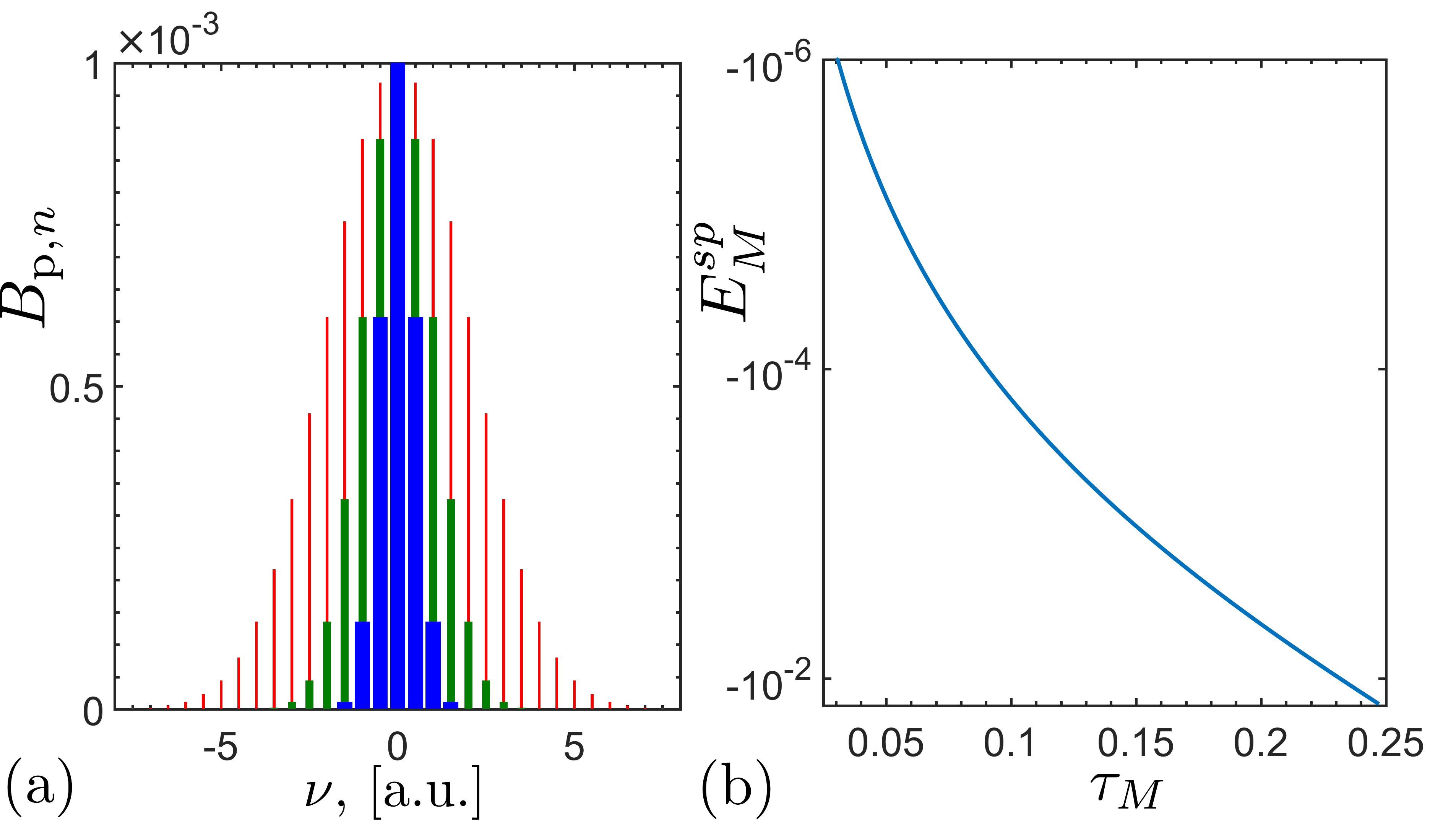}
\caption{ (a) Various spectral densities of a noiseless QOFC with
$M=200$ twin beams with the mean photon numbers of pairs obeying
the Gaussian distribution $B_{\rm
p,n}\equiv10^{-3}\exp(-\nu_n^2/2\sigma^2)$,  with $\sigma=2$ (red
narrow bars), $\sigma=1$ (green thicker bars), and $\sigma=0.5$
(blue thickest bars).  Each tooth in panel (a) represents a twin
beam with two spatially-separated modes of the same frequency.
(b)  NI $E_M^{\sp}$ for noiseless QOFC, according to
Eq.~(\ref{Esp_pure}), versus the nonclassicality depth $\tau_M$,
defined in Eq.~(\ref{tauM}), for different spectra  $B_{\rm p,n}$
shown in panel (a) but with $\sigma$ in the range
$\sigma\in[0,5]$.  The larger is the spectral density of the QOFC,
the larger is the nonclassicality depth $\tau_M$, and, thus, the
more negative  is $E^{\sp}_M$.}\label{fig3}
\end{figure}
For a multimode spontaneous PDC process, the NI $E_M$, given in
Eq.~(\ref{E_M}), attains the form
\begin{equation}\label{EMsp_gen1}  
E_M^{\sp}=\sum\limits_{n=1}^{M}B_{{\rm
s},n}^2\sum\limits_{n=1}^{M}B_{{\rm
i},n}^2-\left(\sum\limits_{n=1}^{M}|D_{{\rm si},n}|^2\right)^2.
\end{equation}
For the symmetric case, when $\sum\limits_{n=1}^{M}B_{{\rm
s},n}^2=\sum\limits_{n=1}^{M}B_{{\rm i},n}^2$,
Eq.~(\ref{EMsp_gen1}) reduces to
\begin{equation}\label{EMsp_gen}  
E_M^{\sp}=\sum\limits_{n=1}^{M}\left(B_{{\rm s},n}^2-|D_{{\rm
si},n}|^2\right) \sum\limits_{n=1}^{M}\left(B_{{\rm
s},n}^2+|D_{{\rm si},n}|^2\right).
\end{equation}
It is clearly seen that the first sum in Eq.~(\ref{EMsp_gen}) is
the sum of the Fourier determinants of the normally-ordered
characteristic functions of each two-mode twin beam, which is  in
analogy to Eq.~(\ref{En}). For the symmetric case,  $E_M^{\sp}$
becomes the sum of the nonclassicality monotones of each two-mode
twin beam. If the $n$th two-mode twin beam is entangled, then it
contributes to the total negativity of  $E_M^{\sp}$. Hence, the
larger is the number of entangled two-mode twin beams in the
system, the larger is the negativity of $E_M^{\sp}$. The number
$M$ of modes  serves as a coherent amplifier  for the
entanglement detection of $E_M^{\sp}$, also due to the last
positive sum in Eq.~(\ref{EMsp_gen}).

For pure multimode twin beams, $E_M^{\sp}$ in Eq.~(\ref{EMsp_gen})
can be written as follows
\begin{equation}\label{Esp_pure}  
E_M^{\sp}=-\sum\limits_{n=1}^MB_{{\rm
p},n}\left(\sum\limits_{n=1}^MB_{{\rm p},n}(2B_{{\rm
p},n}+1)\right).
\end{equation}
Figure~\ref{fig3} shows the dependence of $E_M^{\sp}$ on the Lee's
nonclassicality depth $\tau_M$, defined in Eq.~(\ref{tauM}), for
different spectral distributions of the QOFC displayed in
Fig.~\ref{fig3}(a). Therefore, the larger is the spectral energy
of the QOFC, i.e., the larger is the number of the two-mode twin
beams, the larger is the nonclassicality depth $\tau_M$, and, as a
result, the larger is the negativity of $E_M^{\sp}$.


\subsubsection{Stimulated PDC}
\begin{figure} [t!]
\includegraphics[width=0.4\textwidth]{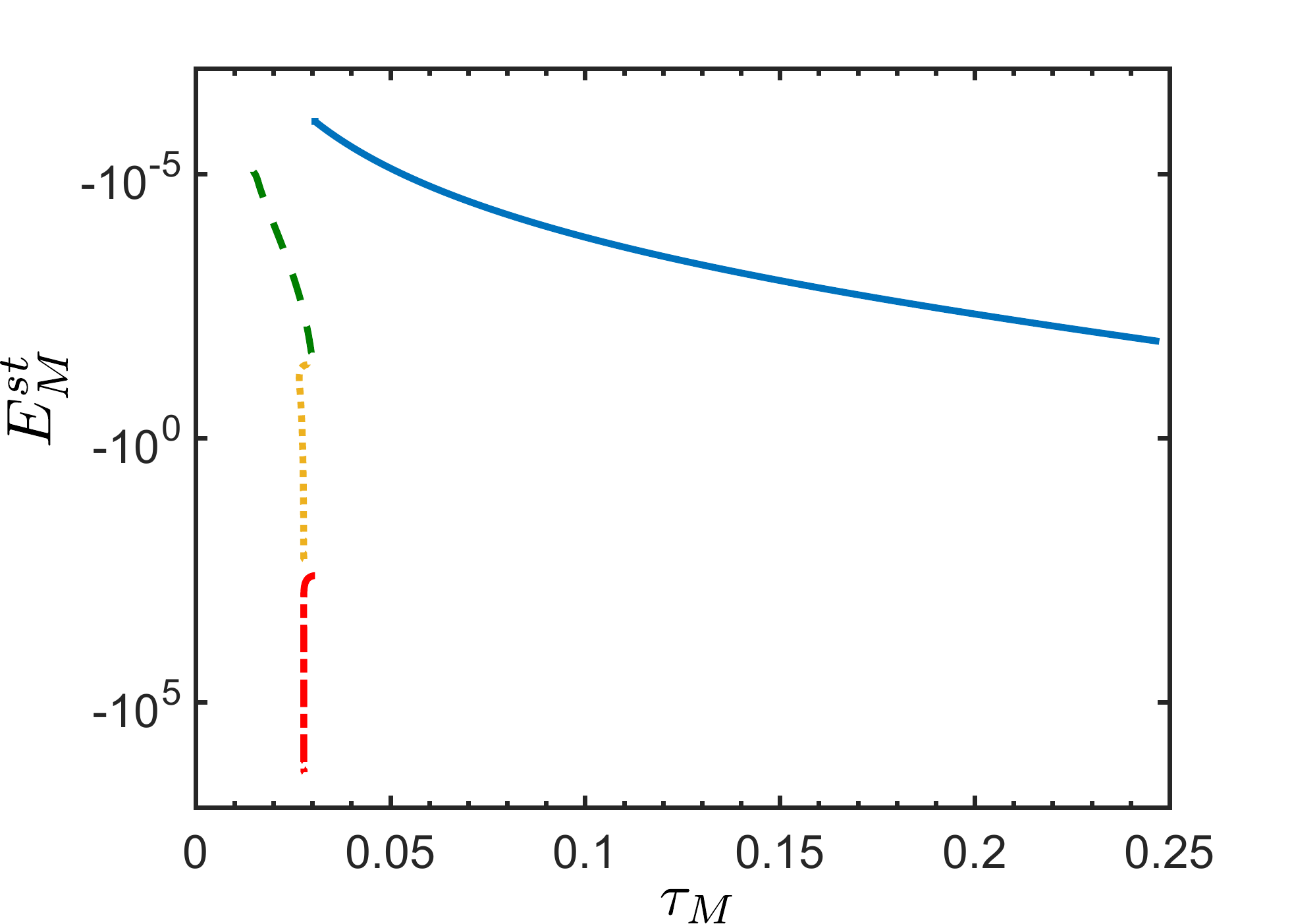}
\caption{Nonclassicality identifier $E_M^{\st}$  versus
nonclassicality depth $\tau_M$ for a given bipartition of the
stimulated noiseless QOFC, where both signal and idler arms
contain $M=200$ modes, for different QOFC spectra at $B_{\rm
p,n}\equiv10^{-3}\exp(-\nu_n^2/2\sigma^2)$, $\sigma\in[0,5]$. The
twin beams are stimulated only in the signal modes. The coherent
stimulating field $\xi_{\rm s}$  in the signal arm is set to:
$\xi_{\rm s}=0$ (blue solid curve),  $\xi_{\rm s}=1$ (green dashed
curve), $\xi_{\rm s}=10$ (orange dotted curve), and $\xi_{\rm
s}=100$ (red dash-dotted curve). The real spectra of the
stimulating coherent field that stimulates the $n$th signal mode
is $|\xi_{{\rm s},n}|\equiv|\xi_{{\rm
s}}|\exp(-\nu_n^2/2\sigma^2)$. With the increasing 
intensities of the stimulating fields, the sensitivity to noise of
$E^{\st}_M$ also increases. Moreover, for very large stimulating
fields, the amount of noise $\tau_M$, needed to make $E^{\st}_M$
positive, becomes independent of the intensity of the coherent
field. }\label{fig4}
\end{figure}
Now, we consider stimulated PDC, when each $n$th signal beam is
stimulated in the signal arm by a coherent field $\xi_{{\rm
s},n}$. Then, the NI $E_M$, in Eq.~(\ref{E_M}), for a bipartite
$M$-mode twin beam state is
\begin{equation}\label{Est_M1}  
E_M^{\st}=E_M^{\sp}-\sum\limits_{n=1}^M|\xi_{{\rm s},n}|^2f_n,
\end{equation}
where $E_M^{\sp}$ is given in Eq.~(\ref{EMsp_gen1}), $f_n$ is a
function of both number of the modes $M$ and elements of
covariance matrix $\boldsymbol{A_{\cal N}}$ of the multimode QOFC.
Whenever each two-mode twin beam of a given QOFC is entangled,
then $f_n\geq0$.  Meaning that, in this case,  stimulating fields
improve the performance of $E_M^{\st}$.

As in the two-mode case, this stimulation leads to the enhancement
of the NI $E^{\st}_M$ (see Fig.~\ref{fig4}). At the same time, as
Fig.~\ref{fig4} shows, $E_M^{\st}$ becomes very sensitive to
noise. Namely, by increasing the  intensities of the stimulating
fields,  $E_M^{\st}$ becomes more negative, but at the expense of
losing the tolerance to larger noise.

We note that, although the application of the NI $E_M$ in
Eq.~(\ref{E_M}) to the multimode twin beam seems straightforward,
in practical situations,  to  separate the signal and idler modes
might be difficult. Thus, the following problem arises: How to
perform an appropriate bipartition that  $E_M$ can detect
conclusively the modes entanglement. In this case, one needs to
implement all possible bipartitions for $E_M$ to reveal the
maximal total entanglement of the multimode twin beam state.


\section{Entanglement identification of QOFC with spatially overlapping  frequency modes}
In this section, we discuss another type of a QOFC, namely, when
the signal mode of a twin beam generated by the $n$th tooth of the
COFC pump  spatially overlaps  with the signal or idler modes of
the other twin beams produced by different or the same OFC teeth.
As a result, one cannot simply divide such QOFC into a set of
independent two-mode twin beams, as it was the case discussed in
Sec.~III.

Now, we consider the following interaction Hamiltonian
\begin{equation}\label{HOFC2}  
\hat H = -\hbar g\sum\limits_{k_s,k_i}\hat a_{k_s}\hat a_{k_i} +
{\rm h.c.},
\end{equation}
where we assume that the coupling strength $g$ for each generated
entangled pair is the same and  real. As Eq.~(\ref{HOFC2})
implies, any spatial frequency mode $k_s$ is equally coupled to
various spatial frequency modes $k_i$. Meaning that a given $k_s$
mode can contain photons that are simultaneously entangled to
different  modes $k_i$.

For this case, when the Hamiltonian in Eq.~(\ref{HOFC2}) contains
$N$ different spatial frequency modes, the evolution $2N\times 2N$
matrix, in Eq.~(\ref{HE2}), takes the form
$\boldsymbol{M}=gL_1\otimes L_2$, where
\begin{equation}  
L_1=\begin{pmatrix}
{0} & {1} \\
{-1} & {0}
\end{pmatrix},\quad L_2=\begin{pmatrix}
0 & 1 & \dots & 1 \\
1 & 0 & \dots & 1 \\
\vdots & \dots & \ddots & \vdots \\
1 & \dots & 1 & 0
\end{pmatrix},
\end{equation}
and $L_2$ is a $N\times N$  hollow matrix of ones, i.e., all its
elements equal one, except the main diagonal elements which are
zero.

The elements of the symmetric exponential matrix
$\boldsymbol{S}=\exp\left(i\boldsymbol{M}t\right)$, in
Eq.~(\ref{AS}), after straightforward but some algebra, can be
found as follows
\begin{eqnarray}\label{S}  
S_{j,j}&=&\frac{1}{2N}\Big(\cosh[(N-1)gt]+(N-1)\cosh [gt]\Big), \nonumber \\
S_{j,j+1}&=&\frac{i}{2N}\Big(\sinh[(N-1)gt]-(N-1)\sinh [gt]\Big), \nonumber \\
S_{j,2k+1}&=&\frac{1}{2N}\Big(\cosh[(N-1)gt]-(N-1)\cosh [gt]\Big), \nonumber \\
S_{j,2k+2}&=&\frac{i}{2N}\Big(\sinh[(N-1)gt]+(N-1)\sinh [gt]\Big), \nonumber \\
\end{eqnarray}
for $j=1,\dots,N$, and $k=1,\dots,N-1$. Having  the matrix
$\boldsymbol{S}$, we can immediately obtain the normally-ordered
covariance matrix $\boldsymbol{A_{\cal N}}$ in
Eq.~(\ref{multiCM}). Thus, by combining  Eqs.~(\ref{multiCM}) and
(\ref{S}), we obtain the elements of the matrix
$\boldsymbol{A_{\cal N}}$, which read as follows
\begin{eqnarray}\label{P2}  
B_{{\rm p},j} &=& \frac{1}{2N}\Big(\cosh[2(N-1)gt]+(N-1)\cosh[2gt]\Big)-\frac{1}{2}, \nonumber \\
C_j &=& \frac{i}{2N}\Big(\sinh[2(N-1)gt]-(N-1)\sinh[2gt]\Big), \nonumber \\
D_{jk} &=& \frac{i}{2N}\Big(\sinh[2(N-1)gt]+(N-1)\sinh[2gt]\Big), \nonumber \\
\bar D_{jk} &=& \frac{1}{N}\Big(\sinh^2[(N-1)gt]-\sinh^2[gt]\Big),
\end{eqnarray}
for $j,k=1,\dots,N$. Since the parameter $B_{{\rm p},j}$ does not
account for mean photon-numbers of pairs anymore, as it was in
Sec.~III, we will call it simply as the mean photon number of
vacuum fluctuations of the spatial frequency mode $j$. Considering
the stimulation process, where each frequency mode of the QOFC is
seeded by a coherent field $\xi_j$, the dynamics of these
stimulating fields obeys Eq.~(\ref{XiS}) with the matrix
$\boldsymbol{S}$, given in Eq.~(\ref{S}).

\subsection{Two-mode entanglement}
\subsubsection{Spontaneous PDC}
For two-mode entanglement of the  QOFC generated in spontaneous
PDC with spatially overlapped  frequency modes, the NI $E$, after
applying Eq.~(\ref{E2}), reads as follows
\begin{equation}\label{Esp2}   
E^{\sp}_{jk} =
\left(B_j^2+|C_j|^2\right)\left(B_k^2+|C_k|^2\right)-\left(|D_{jk}|^2+|\bar
D_{jk}|^2\right)^2,
\end{equation}
where $B_j=B_{{\rm p},j}+\langle n_j\rangle$ with the mean thermal
noise photon-number $\langle n_j\rangle$ in the $j$th mode,  and
$B_{{\rm p},j}$, $C_j$, $D_{jk}$, and $\bar D_{jk}$ are given in
Eq.~(\ref{P2}). For simplicity, we will drop the subscripts in
Eq.~(\ref{Esp2}), as we are interested only in two-mode states.
\begin{figure} 
\includegraphics[width=0.5\textwidth]{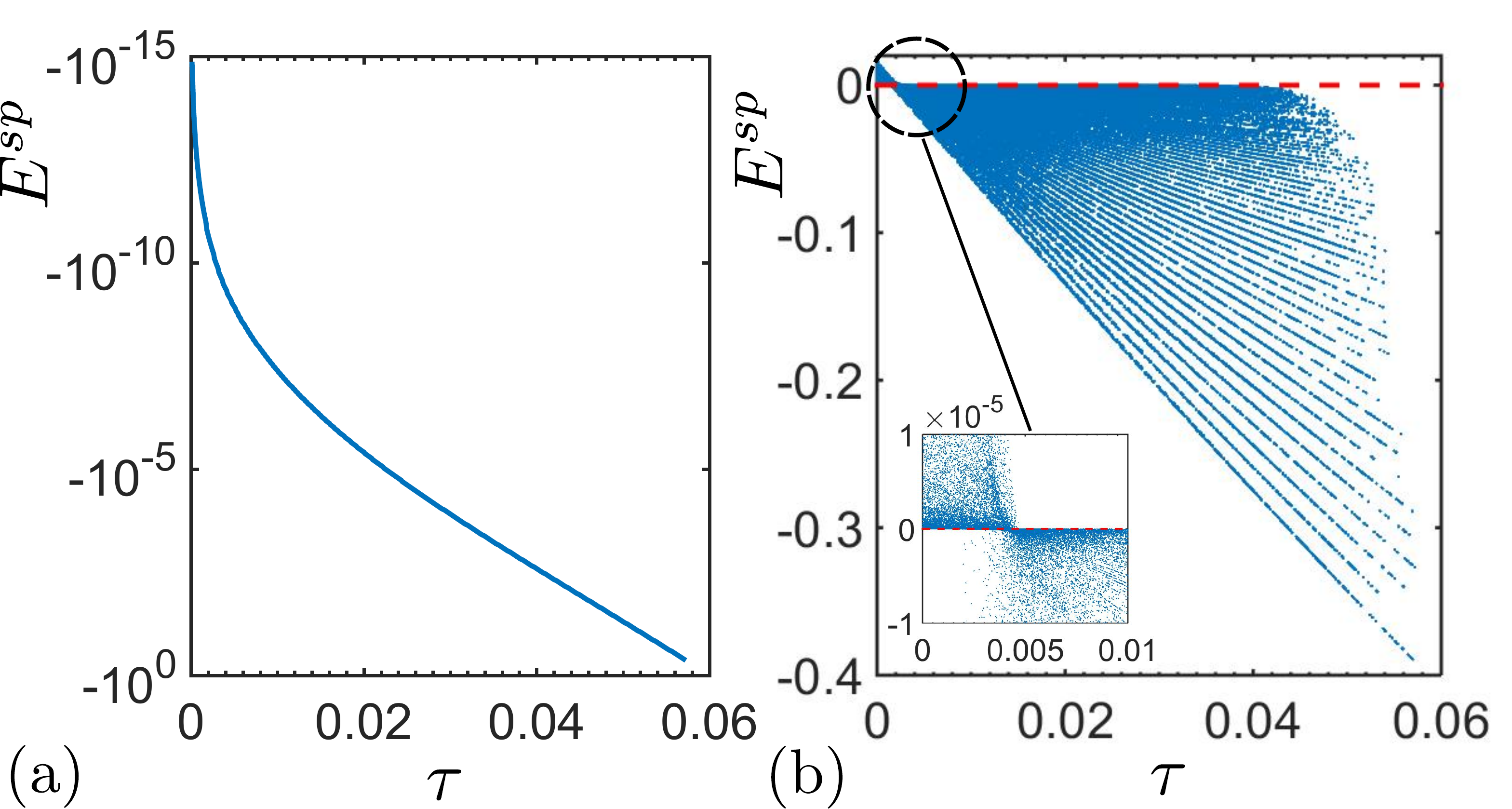}
\caption{(a) Nonclassicality identifier $E^{\sp}$  versus
nonclassicality depth $\tau$ for any two spatial frequency modes
of the noiseless QOFC without stimulation. (b) The same as in
panel (a) but for the mixed two-mode state. The total number of
spatial frequency modes of the generated QOFC is $N=100$, and the
mean photon number of vacuum fluctuations and thermal noise photon
number in each spatial frequency mode is $B_{{\rm p},j},\langle
n_j\rangle\in[0,1]$, respectively. In general,  $E^{\sp}$ is not a
monotone of $\tau$. Nevertheless, whenever $\tau>0.5/N$, $E^{\sp}$
is always a monotone of $\tau$. }\label{fig5}
\end{figure}

For noiseless QOFC, there is one-to-one correspondence between
the NI $E^{\sp}$ and the Lee's nonclassicality depth $\tau$  [see
Fig.~\ref{fig5}(a)]. The latter means that with the increasing
entanglement between two given modes of the QOFC, the negativity
of $E^{\sp}$ also increases.

In general, $E^{\sp}$ in Eq.~(\ref{Esp2}) fails to detect
entanglement between two different spatial frequency modes for
noisy QOFC. Namely, as Fig.~\ref{fig5}(b) indicates, there is a
region of nonclassicality and entanglement, where the NI $E^{\sp}$
is positive. Nevertheless, as our numerical findings show, for a
two-mode state with large nonclassicality, i.e., with large values
of $\tau$, the NI $E^{\sp}$ is always a monotone of $\tau$.
Moreover, for large number of modes, $N\gg1$, generated in QOFC,
there is a bound for $\tau$. Namely, whenever $\tau>0.5/N$,  the
NI $E^{\sp}$ is always negative [see Fig.~\ref{fig5}(b)]. In other
words, with the increasing $N$  number of the modes in the QOFC,
the NI $E^{\sp}$ tends to become a genuine monotone of $\tau$. We
note that the value $\tau=0.5$ is a maximal value of the
nonclassicality depth, which can be reached by a Gaussian state.
\subsubsection{Stimulated PDC}
\begin{figure} 
\includegraphics[width=0.4\textwidth]{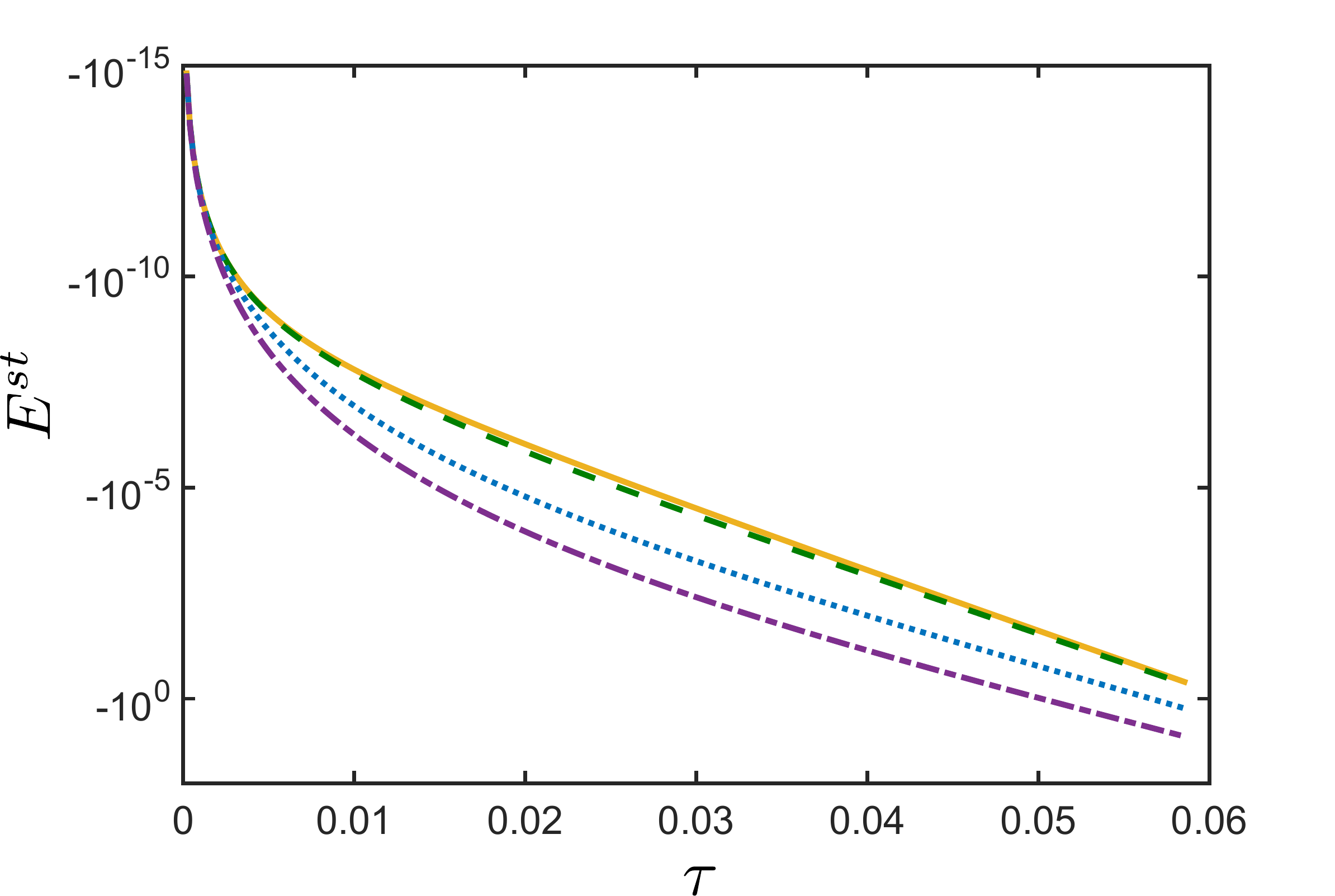}
\caption{Nonclassicality identifier $E^{\st}$ versus
nonclassicality depth $\tau$ for any two spatial frequency modes
of the stimulated noiseless QOFC. The stimulation is applied only
in one spatial frequency mode that does not belong to the given
two-mode state. The amplitude of the stimulating field   is:
$\xi=0$ (yellow solid curve),  $\xi=10$ (green dashed curve),
$\xi=50$ (blue dotted curve),  $\xi=100$ (violet dash-dotted
curve). The total number of spatial frequency modes of the
generated QOFC is $N=100$, and the mean photon number of vacuum
fluctuations  in each spatial frequency mode is $B_{{\rm
p},j}\in[0,1]$. The NI $E^{\st}$ is independent on the phase of
$\xi$. For a given value of $\tau$,  $E^{\st}$ exhibits larger
negative values for larger stimulating-field
amplitudes.}\label{fig6}
\end{figure}
For the stimulated QOFC, the variances $\langle\Delta W_j^m\Delta
W_k^n\rangle_{\cal N}$ of the integrated intensity moments,
defined  in Eq.~(\ref{MOM}), read as follows
\begin{eqnarray}\label{MOM2} 
\langle\Delta W_j^2\rangle_{\cal N}&=&B_j^2+|C_j|^2+2B_j^2|\xi_j|^2+2{\rm Re}\left[C_j{\xi_j^*}^2\right], \nonumber \\
\langle\Delta W_j\Delta W_k\rangle_{\cal N}& = &|D_{jk}|^2+|\bar D_{jk}|^2 \nonumber \\
&&+2{\rm Re}\left[D_{jk}{\xi_j^*}\xi_k^*\right]+2{\rm
Re}\left[\bar D_{jk}{\xi_j}\xi_k^*\right].
\end{eqnarray}
For noiseless QOFC, even when seeding either the $k$th mode that
does not belong to a given two-mode state, the negativity of the
NI $E^{\st}$ increases with the increasing seeding field $\xi_k$
(see Fig.~\ref{fig6}). Moreover, $E^{\st}$ is independent on the
phase of the stimulating signal field $\xi_k$.

\subsection{Multimode bipartite entanglement}
For a bipartite state that contains $M$ modes in both signal and
idler arms, the applied $E_M$, given in Eq.~(\ref{E2}), takes the
following form
\begin{eqnarray}\label{E_M2}  
E_M=&&\left<\left(\sum\limits_{j=1}^M\Delta W_{1,j}\right)^2\right>_{\cal N}\left<\left(\sum\limits_{j=1}^M\Delta W_{2,j}\right)^2\right>_{\cal N} \nonumber \\
&&-\left<\sum\limits_{j=1}^M\Delta
W_{1,j}\sum\limits_{k=1}^M\Delta W_{2,k}\right>^2_{\cal N}.
\end{eqnarray}
We note that, compared to Eq.~(\ref{E_M}),  Eq.~(\ref{E_M2}) has a
more complicated form due to the simultaneous presence of auto-
and cross-correlations in both arms denoted as $W_1$ and $W_2$.
Since each of those arms contains  $M$ modes which are also
entangled. In other words, each term in Eq.~(\ref{E_M2}) consists
of a sum of different single- and two-mode integrated intensity
moments, given in Eq.~(\ref{MOM2}).
\begin{figure} 
\includegraphics[width=0.45\textwidth]{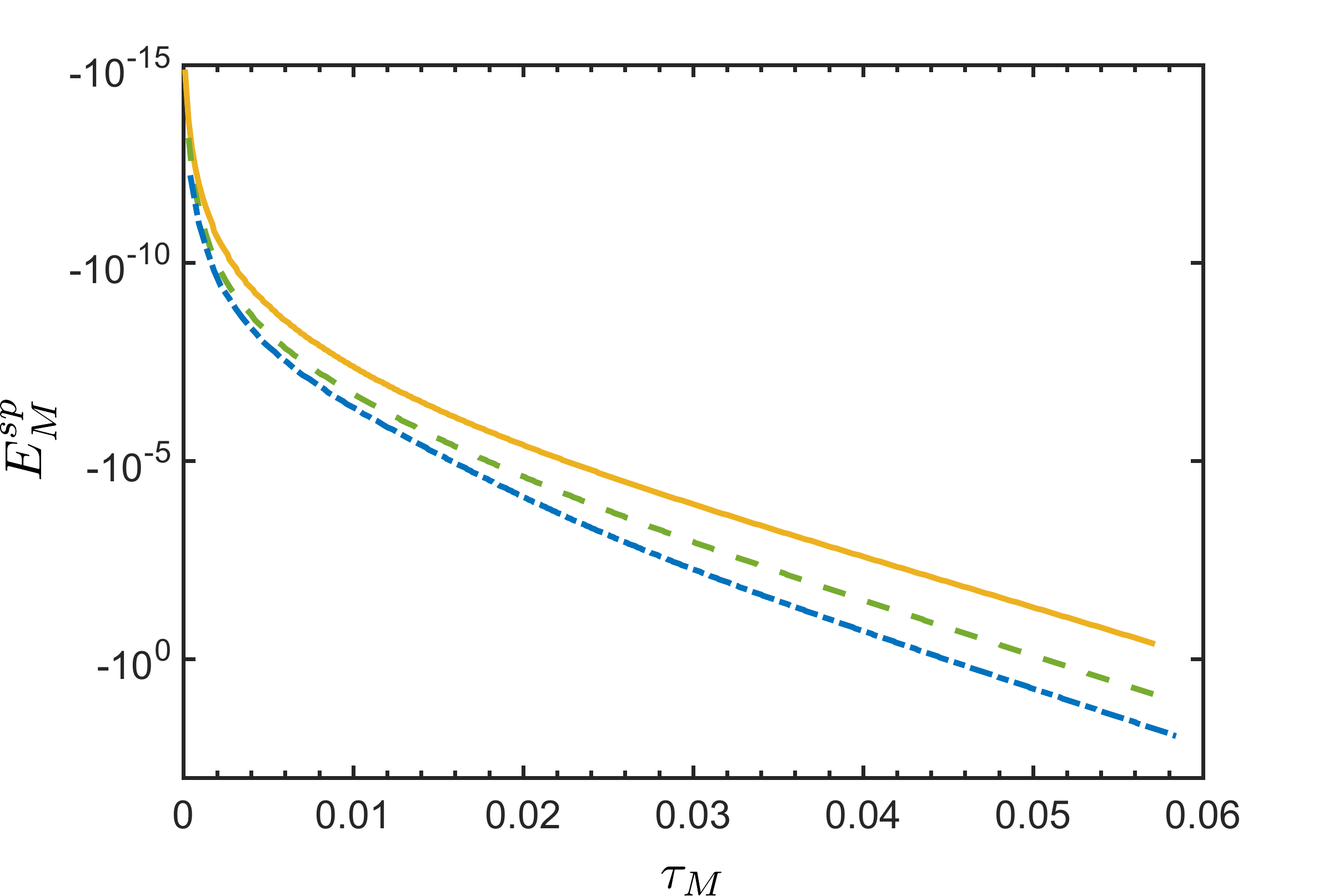}
\caption{Nonclassicality identifier  $E^{\sp}_M$ versus
nonclassicality depth $\tau_M$ for a certain bipartition of the
fields of a noiseless QOFC, where each part contains the following
number of spatially-frequency modes:  $M=1$ (yellow solid curve),
$M=3$ (green dashed curve), and $M=6$ (blue dash-dotted curve).
The total number of spatial frequency modes of the generated QOFC
is $N=100$, and the mean photon number of vacuum fluctuations in
each spatial frequency mode is $B_{{\rm p},j}\in[0,1]$. The NI
$E^{\st}_M$ displays a larger negativity when one includes more
spatial frequency modes in a given bipartition.}\label{fig7}
\end{figure}
\begin{figure} 
\includegraphics[width=0.45\textwidth]{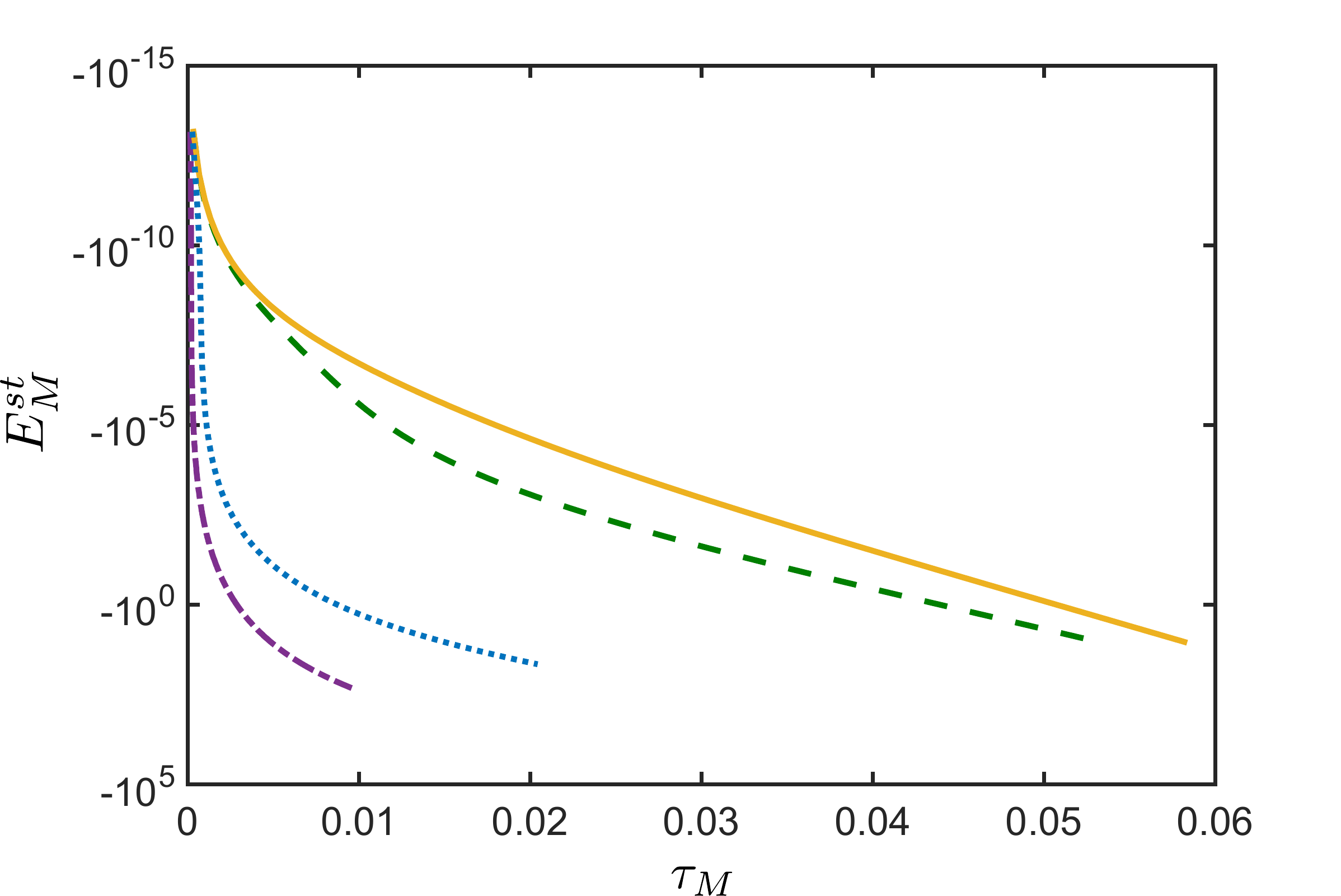}
\caption{Nonclassicality identifier $E^{\st}_M$ versus
nonclassicality depth $\tau_M$ for a bipartition where each part
contains three spatial frequency modes for the case of the
stimulated noiseless QOFC. The stimulation occurs only in one
spatial frequency mode that does not belong to a given
bipartition. The amplitude $\xi$ of the stimulating field  is:
$\xi=0$  (yellow solid curve),  $\xi=10$ (green dashed curve),
$\xi=50$ (blue dotted curve), $\xi=100$ (violet dash-dotted
curve). The total number of spatial frequency modes of the
generated QOFC is $N=100$, and the mean photon number of vacuum
fluctuations in each  mode is $B_{{\rm p},j}\in[0,1]$.
}\label{fig8}
\end{figure}
For a symmetric system, i.e., when  all the modes are
statistically equivalent, the terms in Eq.~(\ref{E_M2}) can be
simplified as follows
\begin{eqnarray}  
\left<\left(\sum\limits_{j=1}^M\Delta W_{1,j}\right)^2\right>_{\cal N}&&=M\left<\Delta W_{j}^2\right>_{\cal N} \nonumber \\
&&+M(M-1)\langle\Delta W_j\Delta W_k\rangle_{\cal N}, \nonumber \\
\left<\sum\limits_{j=1}^M\Delta W_{1,j}\sum\limits_{k=1}^M\Delta
W_{2,k}\right>_{\cal N}&&=M^2\langle\Delta W_j\Delta
W_k\rangle_{\cal N},
\end{eqnarray}
where $\left<\Delta W_{j}^2\right>_{\cal N}$ and $\langle\Delta
W_j\Delta W_k\rangle_{\cal N}$ are given in Eq.~(\ref{MOM2}).
\subsubsection{Spontaneous PDC}
In the case of the spontaneous  PDC, the negativity of the NI
$E^{\sp}_M$ is increasing with the increasing number $M$ of the
modes  involved in a given bipartition (see Fig.~\ref{fig7}). This
means, that by inserting another pair of the spatial frequency
modes into the bipartition, one boosts the performance of
$E^{\sp}_M$ in the entanglement detection of a given  state.

\subsubsection{Stimulated PDC}
For a stimulated QOFC, the  NI $E^{\st}_M$ again enhances its
sensitivity to detect bipartite entanglement (see
Fig.~\ref{fig8}).  But for larger stimulating fields, the NI
$E^{st}_M$ becomes less resistant to noise (see Fig.~\ref{fig8}).
Note that, as in the two-mode case, in order to boost the
performance of  $E^{\st}_M$, it is not necessary to stimulate the
measured fields. It is already enough to seed only one of all the
$N$ modes of the QOFC, which  does not belong to a given
bipartition, in order to make  $E^{\st}_M$ more negative.

\section{Conclusions}
In this study, we have shown the usefulness of the nonclassicality
identifier, given in Eq.~(\ref{E2}), to detect the bipartite
entanglement of the QOFC generated in both spontaneous PDC and
stimulated PDC processes. This NI is expressed via integrated
second-order intensity moments  of the detected optical fields
which makes it a convenient and powerful tool for the experimental
detection of the entangled modes in QOFCs. We have considered two
different cases where a QOFC was comprised either by spatially
non-overlapping or completely overlapping frequency modes. We have
demonstrated that in both cases the NI displays a good performance
in revealing  bipartite entanglement for noisy QOFC. Most
importantly, with the help of strong stimulating fields, one can
sufficiently increase the efficiency of a given NI to reveal the
entanglement of QOFCs, but at the expense of a higher sensitivity
to thermal noise.

%

\end{document}